\documentclass[]{spie}  

\usepackage{amsmath,amsfonts,amssymb}
\usepackage{graphicx}
\usepackage[colorlinks=true, allcolors=blue]{hyperref}

\title{Performance of the MAGIC stellar intensity interferometer and expansion to MAGIC + CTAO-LST1 stellar intensity interferometer}

\author[a]{Alejo Cifuentes}
\author[b]{V.A. Acciari}
\author[c]{F. Barnes}
\author[d]{G. Chon}
\author[b]{E. Colombo}
\author[a]{J. Cortina}
\author[a]{C. Delgado}
\author[a]{C. Díaz}
\author[e]{M. Fiori}
\author[d]{D. Fink}
\author[a]{T. Hassan}
\author[d]{I. Jiménez Martínez}
\author[a]{I. Jorge}
\author[f,g]{D. Kerszberg}
\author[h]{E. Lyard}
\author[a]{G. Martínez}
\author[d]{R. Mirzoyan}
\author[a]{M. Polo}
\author[h]{N. Produit}
\author[a]{J.J. Rodríguez-Vázquez}
\author[i]{P. Saha}
\author[d]{T. Schweizer}
\author[d]{D. Strom}
\author[h]{R. Walter}
\author[j]{C.W. Wunderlich}
\author[*]{on behalf of the MAGIC collaboration and CTAO LST Project\footnote{* A complete list of the MAGIC Collaboration and CTAO LST Project authors can be found at the end of the proceedings}}

\affil[a]{Centro de Investigaciones Energéticas, Medioambientales y Tecnológicas (CIEMAT), Av.Complutense, 40, Madrid, Spain}
\affil[b]{Instituto de Astrofísica de Canarias (IAC), E-38200 La Laguna, Spain}
\affil[c]{Universidad de La Laguna, E-38200 La Laguna, Spain}
\affil[d]{Max-Plank für Physik, D-85748, Germany}
\affil[e]{Università di Padova, I-35131 Padova, Italy}
\affil[f]{Institut de Física d'Altes Energies (IFAE), E-08193 Bellaterra, Spain}
\affil[g]{Sorbone Université, CNRS/IN2P3, Laboratorie de Physique Nucléaire et de Hautes Energies, LPNHE, F-75005 Paris, France}
\affil[h]{Observatorie Astronomique de Genève, CH-1290 Versoix, Switzerland}
\affil[i]{Physik-Institut, University of Zurich, Switzerland}
\affil[j]{Università di Pisa and INFN Pisa, I-56126 Pisa, Italy}

\authorinfo{Further author information: (Send correspondence to A.C.)\\A.C: E-mail: alejo.cifuentes@ciemat.es}

\begin{document} 
\maketitle

\begin{abstract}
A new generation of optical intensity interferometers are emerging in recent years taking advantage of the existing infrastructure of Imaging Atmospheric Cherenkov Telescopes (IACTs). The MAGIC SII (Stellar Intensity Interferometer) in La Palma, Spain, has been operating since its first successful measurements in 2019 and its current design allows it to operate regularly. The current setup is ready to follow up on bright optical transients, as changing from regular gamma-ray observations to SII mode can be done in a matter of minutes.
A paper studying the system performance, first measurements and future upgrades has been recently published. MAGIC SII's first scientific results are the measurement of the angular size of 22 stars, 13 of which with no previous measurements in the B band.

More recently the Large Sized Telescope prototype from the Cherenkov Telescope Array Observatory (CTAO-LST1) has been upgraded to operate together with MAGIC as a SII, leading to its first correlation measurements at the beginning of 2024.
MAGIC+CTAO-LST1 SII will be further upgraded by adding the remaining CTAO-LSTs at the north site to the system (which are foreseen to be built by the end of 2025).
MAGIC+CTAO-LST1 SII shows a feasible technical solution to extend SII to the whole CTAO. 
\end{abstract}

\keywords{Optical Intensity Interferometry, Imaging Atmospheric Cherenkov Telescopes, CTAO, Instrumentation: Interferometers, Instrumentation: High Angular Resolution, Stars: Imaging, Stars: Fundamental Parameters, Stars: Fast Rotators.}

\section{INTRODUCTION}
\label{sec:intro}  
Hanbury-Brown \& Twiss [\citenum{1956Natur.178.1046H}] successfully proved that at sufficiently short baselines, the spatial correlation measurements of stars exhibit a photon bunching signal for measurements within a defined area on the ground, and that this area depends on the angular size of the star. An SII in Narrabri, Australia was used to measure the diameters of the 32 brightest stars in the Southern Hemisphere [\citenum{1974iiia.book.....H}]. This SII was limited by the signal-to-noise ratio (S/N), calling for more light collection or faster photodetectors and electronics, hard to achieve at the time. The development of SIIs was then stopped for around 40 years. 

In the last decade interest towards intensity interferometry has increased again with new generation instrumentation. Already running IACTs, equipped with large reflectors ($>100 m^2$) and fast acquisition chains ($\sim 1 ns$), have developed intensity interferometry as a second observing mode (see VERITAS [\citenum{2020NatAs...4.1164A}], MAGIC [\citenum{2024MNRAS.529.4387A}] and H.E.S.S [\citenum{2024MNRAS.52712243Z}] first intensity interferometry measurements). VERITAS and H.E.S.S rely on a setup that can be mounted on the focal plane of the IACTs while MAGIC approach tries to reuse as much as possible of the existing hardware.

Similarly to MAGIC, CTAO-LST1 has been upgraded to allow it to operate as an SII jointly with MAGIC, making its first correlation measurement at the beginning of 2024. Once the commission is finished, MAGIC+CTAO-LST1 SII will be the most sensitive SII ever built.

\section{MAGIC SII}
The MAGIC telescopes are a system of two IACTs of 17 m mirror dish diameter, located at
an elevation of 2200 m above sea level, at the Roque de los Muchachos Observatory (ORM) in La Palma,
Canary Islands (Spain). MAGIC provides an integral sensitivity of 0.66 ± 0.03 \% of the Crab
Nebula flux above 220 GeV in 50 h of observation, and allows the measurement of photons in the
energy range from 50 GeV to above 50 TeV [\citenum{2016APh....72...76A}].

In recent years, the MAGIC telescopes have undergone several technical modifications ([\citenum{2020MNRAS.491.1540A}, \citenum{2022SPIE12183E..0CC}, \citenum{2022icrc.confE.693D}])
like the implementation of a digitizer and GPU-based real-time correlator or the automatic deployment of optical filters in front of the photomultiplier tubes (PMTs) used for intensity interferomtry. The filters were installed in the existing target holder. Said filters have mainly the purpose of avoiding damaging the PMTs when pointing to bright stars but we also benefit from the fact that a spectral response with sharp spectral cutoffs improves the sensitivity of the measurement (see Section 2.2 in [\citenum{2022SPIE12183E..0CC}]). A special configuration of the active mirror control (AMC) was also developed to focus all the collected starlight into one or two PMTs. All this modifications allow MAGIC to change from standard Very High Energy (VHE, $E>100~GeV$) observations to SII mode in less than a minute and back to the standard VHE observing mode just as quickly (useful for transients). This design  has proven to be very effective in making MAGIC SII an extremely high duty cycle instrument and since 2021, more than 700 hours of observations have been performed with MAGIC SII. Intensity interferometry observations are regularly performed during Moon breaks, which are periods with bright Moon-light conditions in which IACTs cannot perform sensitive VHE observations, but the intensity interferometer can, thanks to the brightness of the observed stars and the optical filters that greatly reduce the amount of background Moon-light that gets into the PMT. In section \ref{sec:CTAO-LST1 upgrades} we will describe similar upgrades done in CTAO-LST1 that will allow MAGIC+CTAO-LST1 SII to have a high duty cycle too.

\subsection{MAGIC SII's scientific results and performance}
\begin{figure}
    \centering
    \includegraphics[width=0.5\columnwidth]{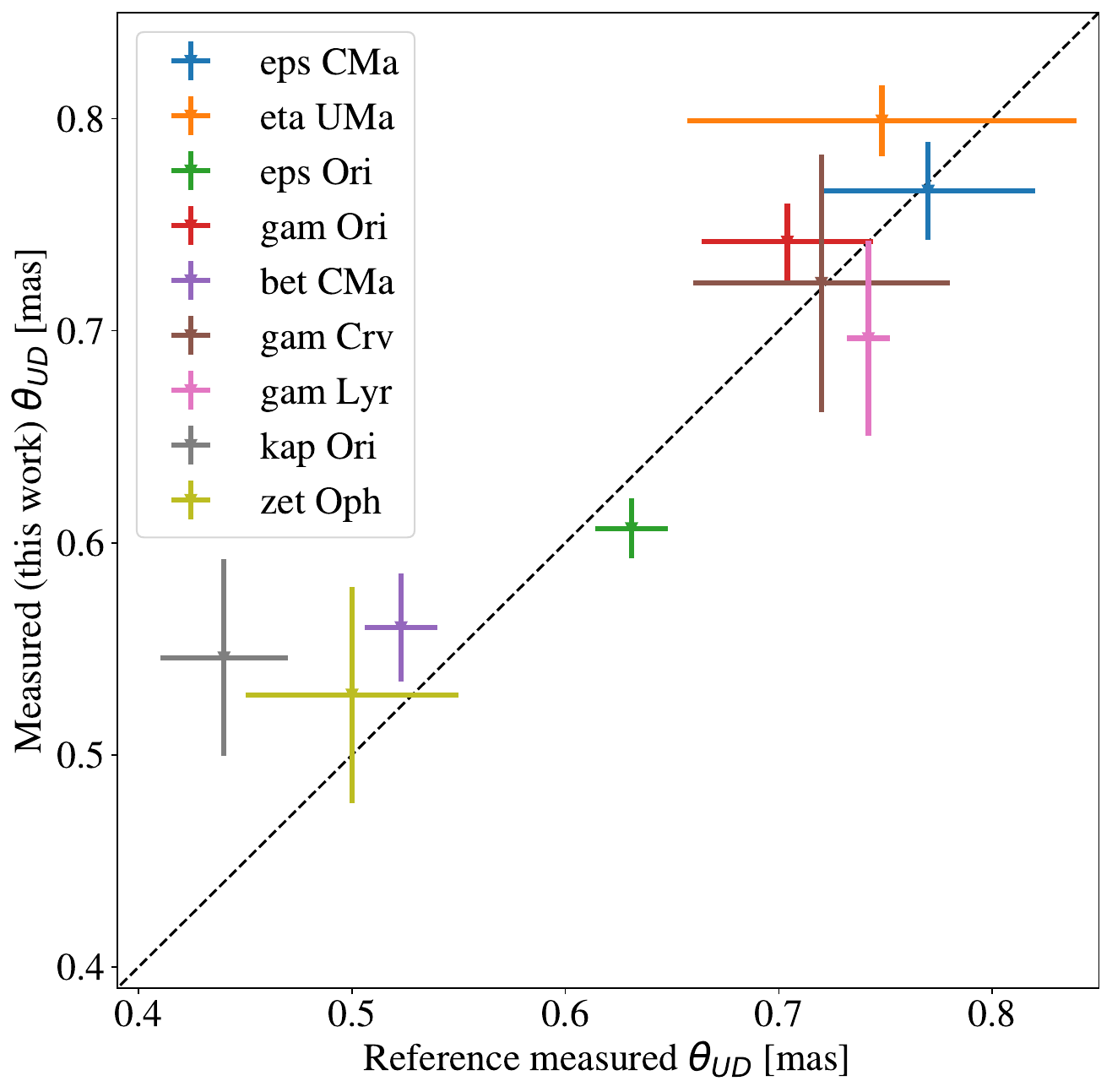}
    \caption{\textit{Direct measurement of the stellar angular diameter of 9 reference
stars, assuming an uniform disk profile, as a function of their reference diameter, also coming from a direct measurement over similar wavelengths (Fig. 9 from [\citenum{2024MNRAS.529.4387A}]). See Table 2 from [\citenum{2024MNRAS.529.4387A}] for their stellar angular diameter assuming a uniform ($\theta_{UD}$) or a limb darkened model ($\theta_{LD}$) , assumed physical parameters
and the source of the reference measurement. Dashed black line shows where
the reference equals the measured diameter values.}}
    \label{fig:refernce_stars}
\end{figure}

\begin{figure}
    \centering
    \includegraphics[width=0.5\columnwidth]{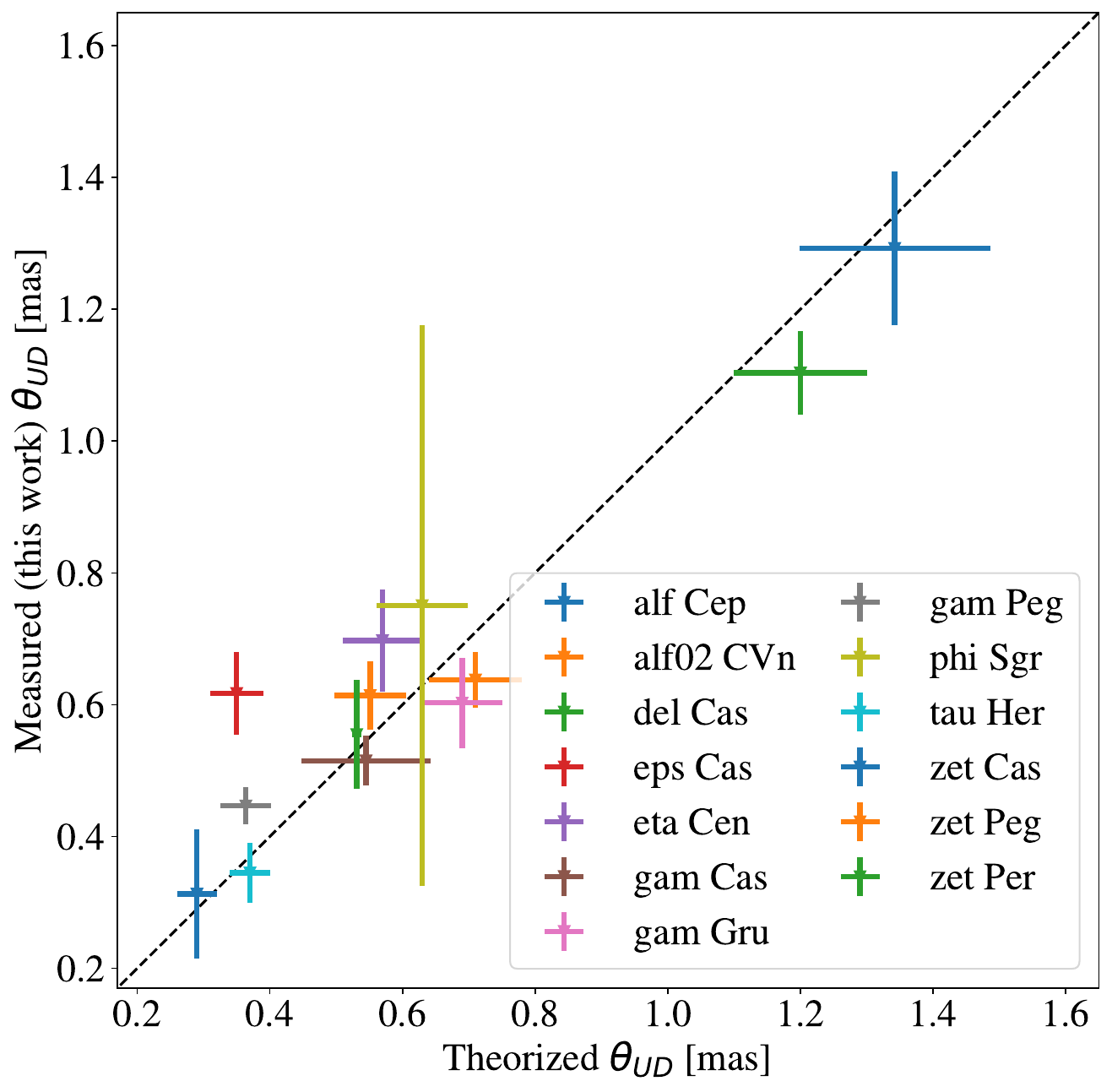}
    \caption{\textit{Direct measurement of stellar angular diameter of 13 stars
measured in the B band for the first time by MAGIC SII, assuming an uniform disk profile, as a function of their expected angular diameter from [\citenum{2014ASPC..485..223B}] (Fig. 10 from [\citenum{2024MNRAS.529.4387A}]). See Table 3 from [\citenum{2024MNRAS.529.4387A}] for
their $\theta_{UD}$ , $\theta_{LD}$ and assumed physical parameters. Note the large uncertainty
associated to phi Sgr is due to the very short observation time acquired
(15 min).}}
    \label{fig:candidate_stars}
\end{figure}

The upgraded MAGIC SII set up was designed, installed and commissioned by 2022. Since then in the period between January and December 2022, 192 hours of data were acquired for 22 targets broadly classified into two categories: 9 \textit{reference} stars (Fig. \ref{fig:refernce_stars})  and 13 \textit{candidate} stars (Fig. \ref{fig:candidate_stars}). A target was considered a reference star if its angular diameter was already measured by other instruments over similar wavelengths (400-440 nm). On the other hand, a target was considered a candidate star if its angular diameter was measured for the first time in the B band.  

The sensitivity of the MAGIC SII was evaluated too, using equation 5.17 from [\citenum{1974iiia.book.....H}] one can calculate the S/N we expect for a correlation signal. The equation, expanded to also account for the effect of the night sky background (NSB) and ignoring extra noise from the readout chain:
\begin{equation}
    S/N = A \cdot \alpha(\lambda_0) \cdot q(\lambda_0) \cdot n(\lambda_0) \cdot |V|^2(\lambda_0 , d) \cdot \sqrt{b_{\nu}} \cdot F^{-1} \cdot \sqrt{T/2} \cdot (1 + \beta)^{-1} \cdot \sigma
    \label{eq:Signal_to_noise_eq}
\end{equation}
Where $A$ is the mirror area, $\alpha(\lambda_0)$ is the quantum efficiency at the peak of the optical passband, $q(\lambda_0)$ the optical efficiency of the rest of the system, $n(\lambda_0)$ is the stellar differential photon flux, $|V|^2(\lambda_0 , d)$ is the squared visibility at the observed wavelength and baseline, $b_{\nu}$ the effective cross-correlation electrical bandwidth, $F$ the excess noise factor of the PMTs, $T$ the observation time, $\beta$ the average background to starlight ratio and $\sigma$ the normalized spectral distribution of the optical passband (see equation 5.6 in [\citenum{1974iiia.book.....H}] for the definition of $\sigma$). More precisely, parameters $A$, $\alpha$, $q(\lambda_0)$, $F$, $\beta$ and $\sigma$ must be understood as the geometric mean between the two telescopes. Furthermore, if we consider an array for identical $N_{telescopes}$, we would have that S/N increases linearly with the square root of the number of telescope pairs. Since $N_{pairs} = N_{telescopes} ( N_{telescopes} - 1) / 2$  we see that N/S increases roughly linearly with the number of identical telescopes. We will come back to this in Section \ref{sec:sicience simulations} and show the expected increase in sensitivity when adding CTAO-LST1 to the system (See Figure \ref{fig:sensitivibity_sim}).

Expected and obtained signal to noise was compared for all the MAGIC SII's acquired data and, when taking into account the effect of the weather conditions, we could see that the measurements matched the expected signal to noise (see Figure 11 from [\citenum{2024MNRAS.529.4387A}]).


The systematic errors affecting the MAGIC SII system were also studied in detail (see [\citenum{2024MNRAS.529.4387A},\citenum{Jimenez-Martinez:2023oN}]). The systematic uncertainties are shown in Table \ref{tab:systematics}. 

\begin{table}[h]
\centering
\begin{tabular}{l|r} 
Systematic effect & Uncertainty \\
\hline
Electronic bandwidth & 0.5\% \\
Optical bandwidth & $<$ 1\% \\
Gain evolution of DC ADC branch & \\
- Seasonal temperature & Negligible \\
- Gain drift after DC jump & 1\% \\
- Long-term degradation & 0.8\% \\
- Deviations from linearity & Negligible \\
Residual electronic noise & Negligible \\
$I_i(\text{NSB})$ subtraction  & 1.5/3\% ($B_{mag} < 3.5 ~/~  > 3.5$) \\
\end{tabular}
\vspace{3mm}
\caption{\textit{Evaluated systematic uncertainties over squared visibility measurements identified to affect the MAGIC SII system (table 5 from [\citenum{2024MNRAS.529.4387A}]).}
}
\label{tab:systematics}
\end{table}

\section{MAGIC + CTAO-LST1 SII}
The LSTs will be the largest IACTs of the CTAO, with a parabolic mirror dish of 23m diameter. LSTs target the lowest energies accesible to CTAO, down to a threshold energy of $\simeq 20~GeV$. LST1 is the first LST built in the CTAO North site, next to the MAGIC telescopes. The first LST1 sky observations took place in December 2018, while regular LST1 observations began in November 2019 [\citenum{2022icrc.confE.872C}].

During 2023, CTAO-LST1 has undergone technical modifications that allowed to perform Intensity Interferometry observations, jointly with MAGIC, for the first time in 2024. MAGIC+CTAO-LST1 SII is now being commissioned and a total of $\sim25h$ of observations have been gathered by the end of May 2024. Observations were mainly done for stars with stellar angular diameter already measured by MAGIC or other instruments, i.e reference stars.

\subsection{CTAO-LST1 upgrades}
\label{sec:CTAO-LST1 upgrades}
Similarly to MAGIC, CTAO-LST1 SII's implementation builds on top of existing hardware, having no impact on regular data taking while allowing for new science. Minimal hardware modifications were done in the CTAO-LST1, including the mounting of optical filters on the existing star imaging target in order to avoid damaging the PMTs when pointing to bright stars and allowing to change between regular gamma-ray observations to SII mode and back to gamma-ray observations in a matter of minutes. One front-end board was redesigned to extract the signal of one PMT from the analog trigger circuit. A dedicated amplifier was added to get the copy of the analog signal without disturbing the normal operation of the trigger. The analog signal copy is sent to a newly installed optical transmitter that transfers this signal through an optical fiber to the same correlator used by MAGIC SII (See section 2.4 in [\citenum{2024MNRAS.529.4387A}]). 



\begin{figure}
    \centering
    \includegraphics[width=0.8\columnwidth, trim={0 3cm 0 0}]{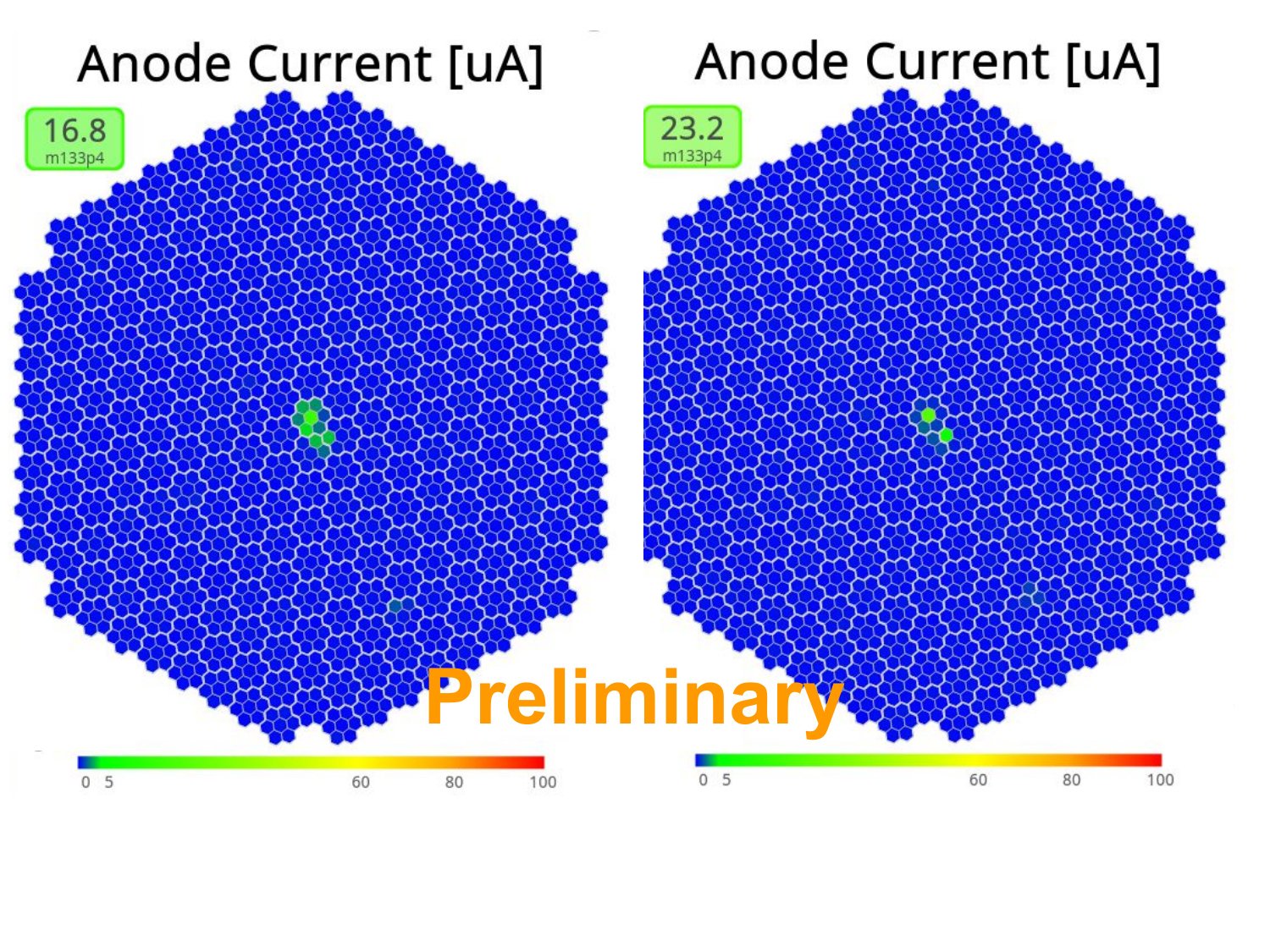}
    \caption{\textit{Comparison of the current measured at the PMTs using the starlight of Mizar (visual double star with a separation of $14.4~arcseconds$), the color code shows the measured anode current at each PMT of the CTAO-LST1 camera. The anode current at the central PMT (the one used for SII observations) is also displayed. On the left panel the standard AMC configuration was used, and the spot of the stars is diffussed among several PMTs around the central one. On the right panel a custom AMC configuration was used, and we can see the light of Mizar focused into the central PMT and a secondary spot for the companion star. The anode current measured in the central PMT increased by a 38\% when using the custom AMC configuration for SII.}}
    \label{fig:AMC_DC}
\end{figure}

In order to achieve the expected sensitivity (Fig. \ref{fig:sensitivibity_sim}) we are also testing an AMC configuration to focus all the starlight into one pixel (see [\citenum{2024MNRAS.529.4387A}, \citenum{2022SPIE12183E..0CC}] for the use of special AMC configurations during MAGIC SII observations). First tests have been done, showing an increase of $\sim38\%$ in the measured anode current (proportional to the collected starlight) of the central PMT when changing from standard to SII AMC configuration (Fig. \ref{fig:AMC_DC}). 


\subsection{MAGIC + CTAO-LST1 SII correlation signal}

A correlation signal was significantly measured in all three baselines of the MAGIC+CTAO-LST1 SII simultaneously for multiple stars up to B$_{\text{mag}} =$ 3.6 and stellar angular diameter of $\sim0.3~mas$. We have mostly observed stars of known diameter in order to calibrate the instrument and understand its performance in a similar way as it was done in the past with MAGIC SII [\citenum{2022SPIE12183E..0CC}, \citenum{2024MNRAS.529.4387A}].

\begin{figure}
    \centering
    \includegraphics[width=0.65\columnwidth, trim={0 3cm 0 0}]{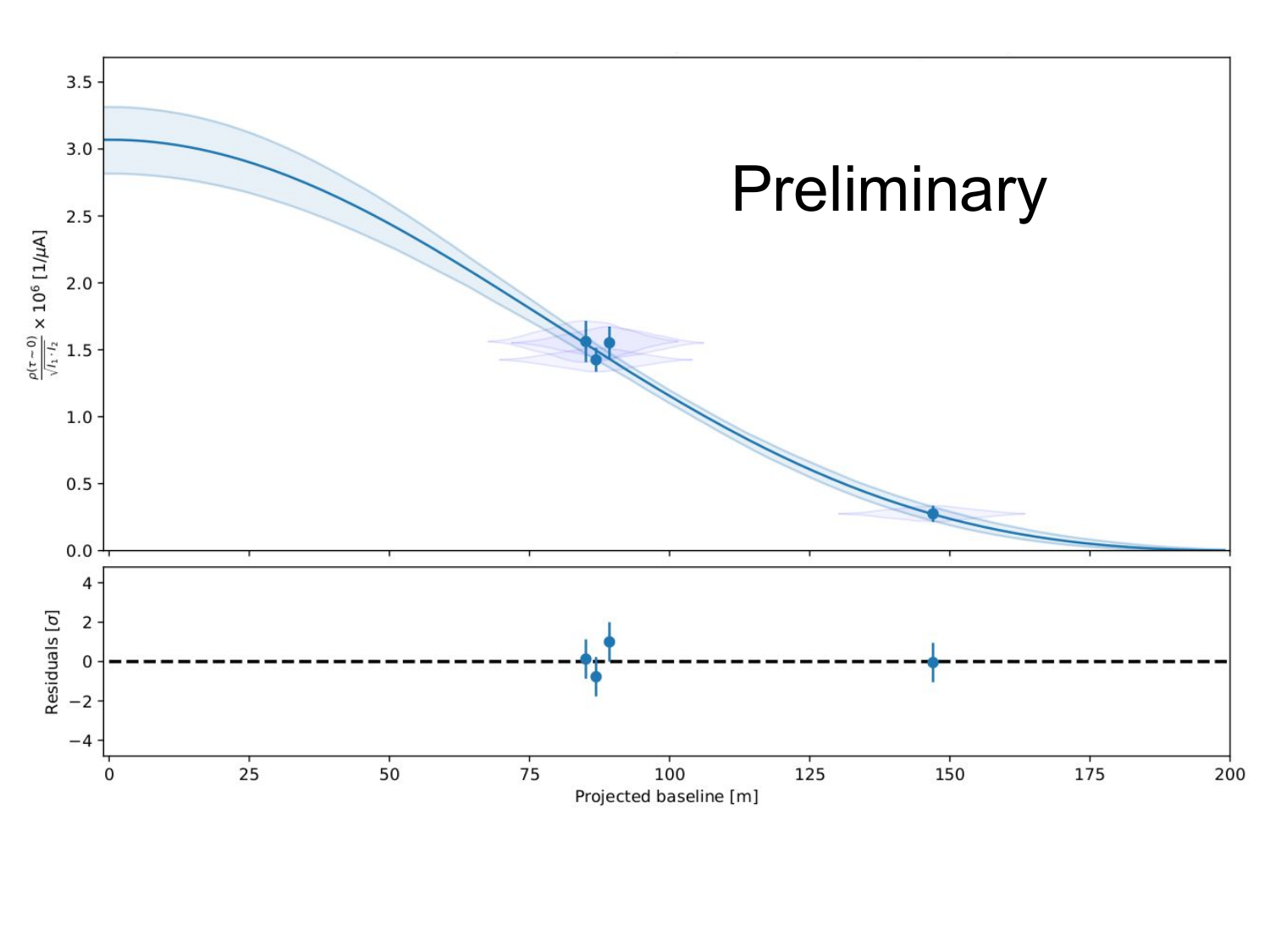}
    \caption{\textit{Preliminary visibility plot for one hour of observations of a $\sim 0.5~ mas$ star. Here only the visibility of the new available baselines is shown. With this measurements we could determine the stellar angular diameter of this target with a $\sim 5\%$ uncertainty.}}
    \label{fig:baseline_vs_visb}
\end{figure}

Due to the increase in sensitivity (See Section \ref{sec:sicience simulations} and Figure \ref{fig:sensitivibity_sim}) and the fact that we have now three baselines instead of one (See Figure 6 from [\citenum{2022icrc.confE.872C}] for the telescopes positions) we can obtain measurements of the stellar angular diameter much faster. In Fig. \ref{fig:baseline_vs_visb} the visibility vs baseline of one hour of observation is shown, where an extended range of baselines is covered simultaneously, allowing for more accurate measurements of angular sizes in less observing time.

\section{Science simulations}
\label{sec:sicience simulations}

With the addition of CTAO-LST1 to the MAGIC SII there are multiple relevant studies that can be done in the near future and mid-term with the addition of the upcoming CTAO-LSTs (expected to be operative at the end of 2025) to the system  (MAGIC+4CTAO-LSTs SII). 

\begin{figure}
    \centering
    \includegraphics[width=0.55\columnwidth]{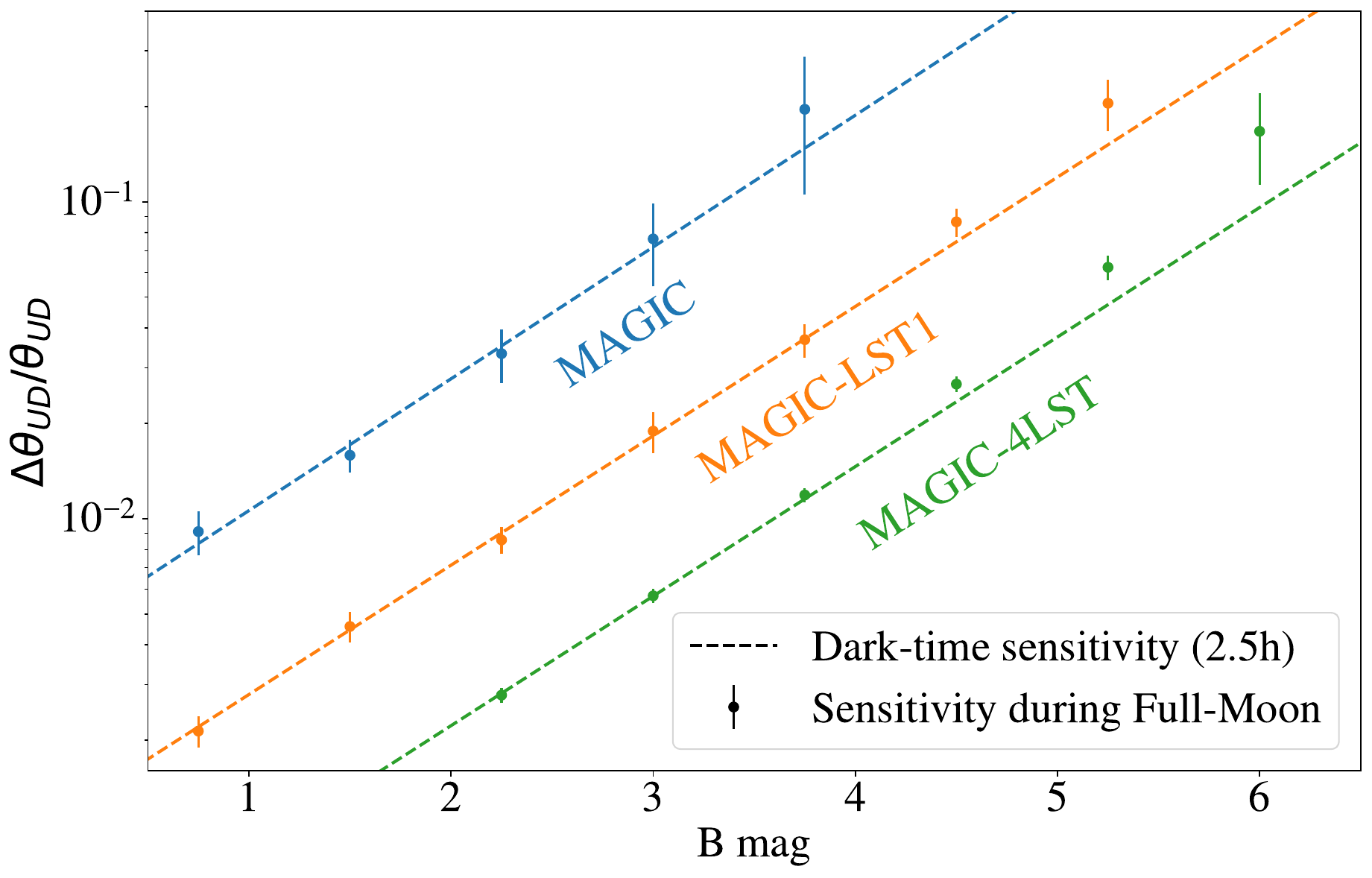}
    \caption{\textit{Relative uncertainty of measured stellar angular diameters as a function of their B magnitude for MAGIC SII (blue), an extension to MAGIC+CTAO-LST1 (orange), and a further extension to MAGIC+4CTAO-LSTs (green). Dashed lines indicate sensitivity during dark time, while solid points estimate the sensitivity under the conservative assumption of observing under very-high NSB illumination levels (both for a 2.5 h observing time). The simulated stellar position and angular diameter is the one from gam Crv  (Fig 13 from [\citenum{2024MNRAS.529.4387A}]).}}
    \label{fig:sensitivibity_sim}
\end{figure}

First of all, adding CTAO-LST1 to the system greatly improves its sensitivity, using Eq. \ref{eq:Signal_to_noise_eq} we can compute the expected signal to noise for MAGIC+CTAO-LST1 SII, which is a factor $\sim 3.6$ better than MAGIC SII alone and the further increase in sesitivity for the MAGIC+4CTAO-LSTs SII, which will be a factor $\sim 11$ more sensitive than MAGIC SII (Fig. \ref{fig:sensitivibity_sim}). This alone increases dramatically the sample of stars that can be observed with MAGIC+CTAO-LST1 SII.

\begin{figure}
    \centering
    \includegraphics[width=0.6\columnwidth]{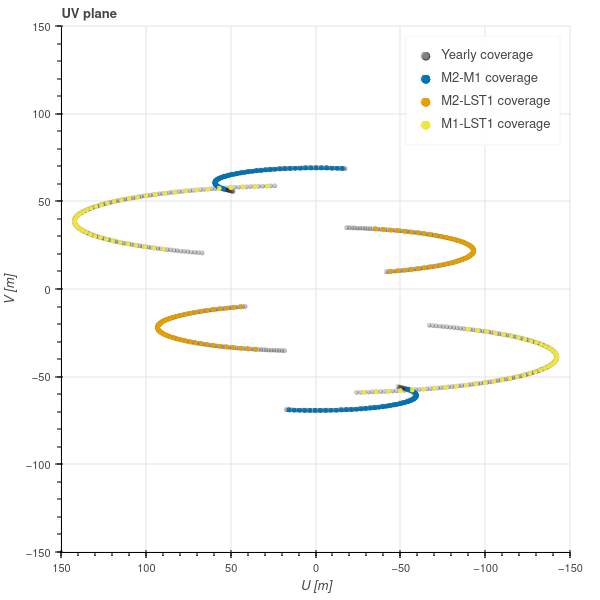}
    \caption{\textit{UV coverage for a star with DEC $\simeq 8 ~deg$ observed with MAGIC+CTAO-LST1 SII. The colored points show the UV plane coverage for a single night for the M2-M1 (blue), M2-LST1 (orange) and M1-LST1 (yellow) baselines. In grey the UV plane coverage of each baseline is extended to the whole year for the same star. Note that with MAGIC SII we could only cover the M2-M1 baseline.}}
    \label{fig:UV_coverage}
\end{figure}

Not only this, but also the angular resolution will improve (larger baselines) as well as the UV plane coverage, since we are observing simultaneously with three baselines now (Fig. \ref{fig:UV_coverage}).

\subsection{Fast rotators: oblateness studies}

\begin{figure}
    \centering
    \includegraphics[width=0.7\columnwidth]{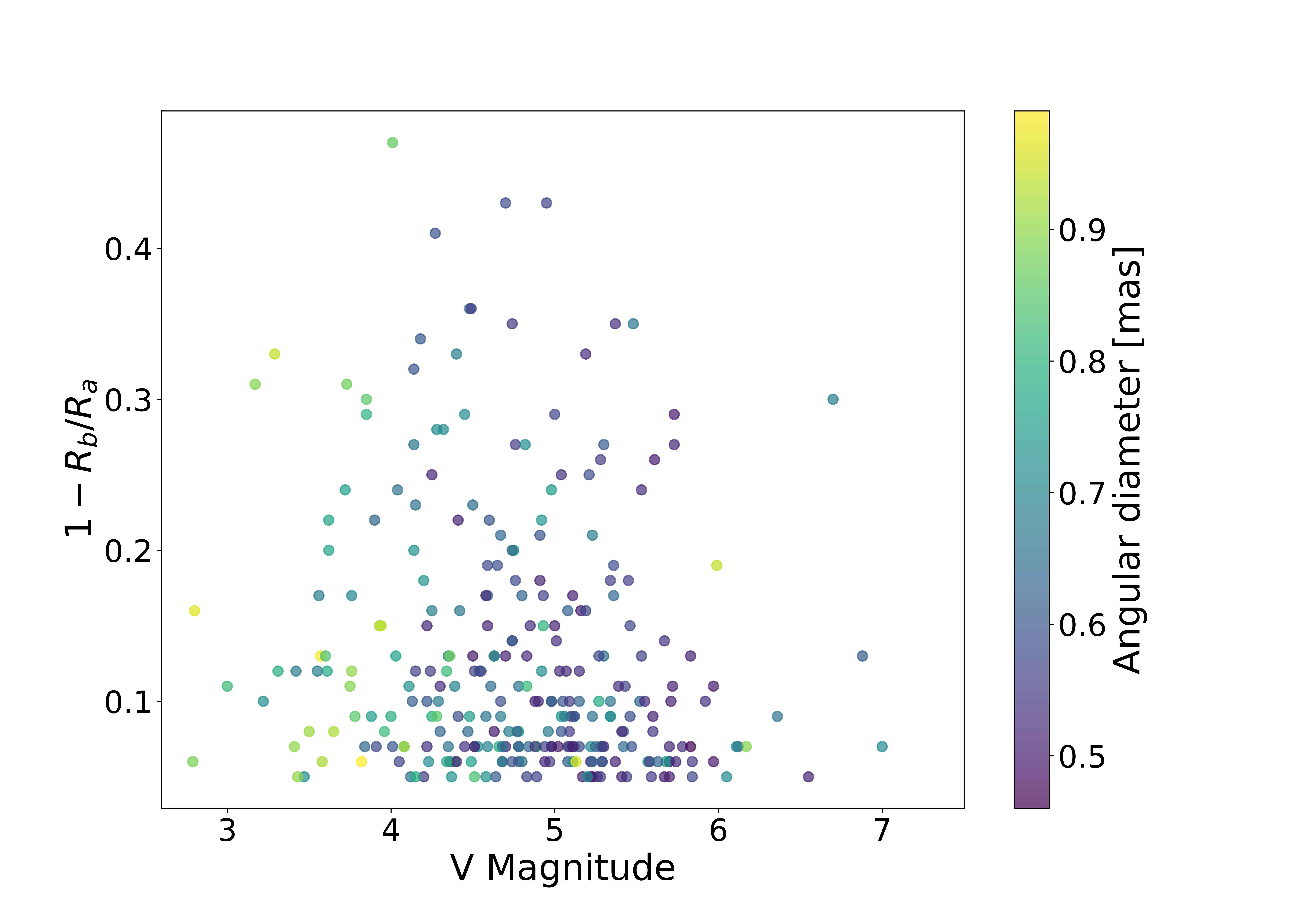}
    \caption{\textit{Targets for oblateness studies (Table 4 in [\citenum{2012A&ARv..20...51V}]). Targets were filtered to stars with angular diameter $< 1 mas$. In the y-axis, $R_b$ and $R_a$ represent the minor and major axis respectively. The color code represents the size of the major axis. The oblateness of the majority of these targets can be already measured with MAGIC+CTAO-LST1 SII.}}
    \label{fig:van_belle}
\end{figure}

Having an improved UV plane coverage allows us to do stellar shape studies, we could measure for the first time the oblateness of multiple fast rotators. There are a lot of known candidates for oblateness studies that have not been measured before (See Table 4 in \citenum{2012A&ARv..20...51V}) (Fig. \ref{fig:van_belle}). 

It is of special interest the case of Be stars, stars with a decretion disk around the equator [\citenum{2013A&ARv..21...69R}]. Instruments capable of resolving the oblateness of Be stars are typically more sensitive to the disk, which is brighter at longer wavelengths (see [\citenum{2013ApJ...768..128T}] and references cited therein). MAGIC+CTA-LST1 SII is, on the other hand, sensitive to the B band and will be able to measure for the first time the stellar shape of multiple Be stars with already measured disk shapes.

\begin{figure}
    \centering
    \includegraphics[width=1\columnwidth]{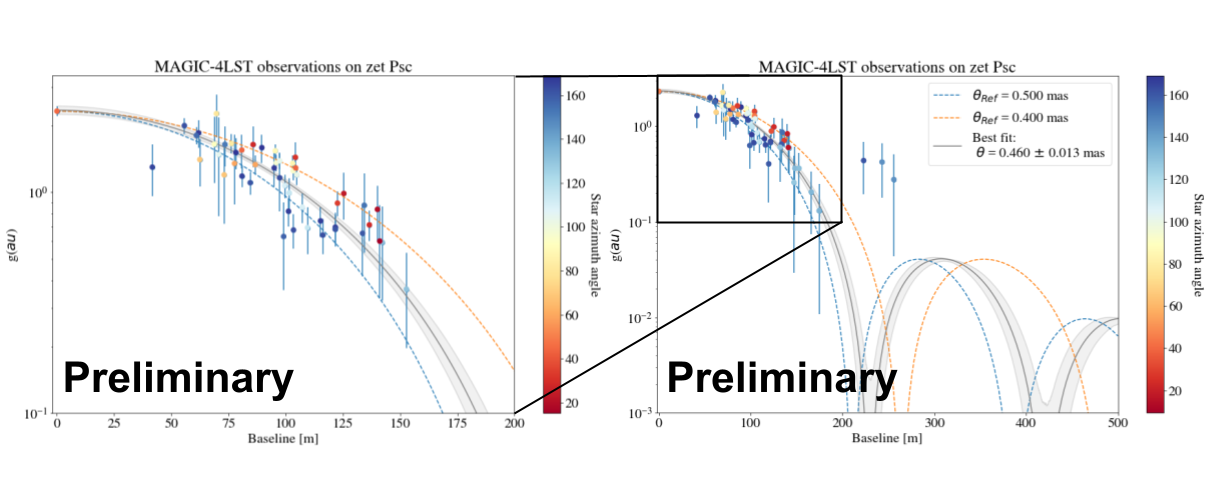}
    \caption{\textit{Simulation of the case of a $B=5.45~mag$ star with an equatorial size of 0.5 mas and 0.25 oblateness observed with MAGIC+4CTAO-LSTs. On the right site simulated observations up to a maximum baseline of $\sim 250~m$ are shown while on the left site, a zoom-in of shorter baselines is shown for greater detail. The color map shows the angle in the UV plane of the simulated observations. The grey line shows a fit of the simulated observations to an uniform disk, while the blue and orange dashed lines show the case of uniform disk models of $0.5~mas$ and $0.4~mas$ respectively. Looking at the zoomed region on the left site, we see that a simple uniform disk model fails to describe the simulated observations, so we instead fitted a uniform ellipse model to the simulated data. In five observations of one hour each we could measure the uniform ellipse model parameters with roughly a $\sim 2\%$ error in the equatorial diameter, $\sim 5~deg$ error in the orientation and $\sim 5\%$ error in the oblateness.}}
    \label{fig:fast_rotator}
\end{figure}

For MAGIC+4CTAO-LSTs SII we have simulated the study of the fast rotator zet Psc (Fig. \ref{fig:fast_rotator}) with $B=5.45~mag$ than could be modelled in a single night with this future expansion of our system.

\subsection{Novae}

\begin{figure}
    \centering
    \includegraphics[width=0.7\columnwidth, trim={0 3cm 0 0}]{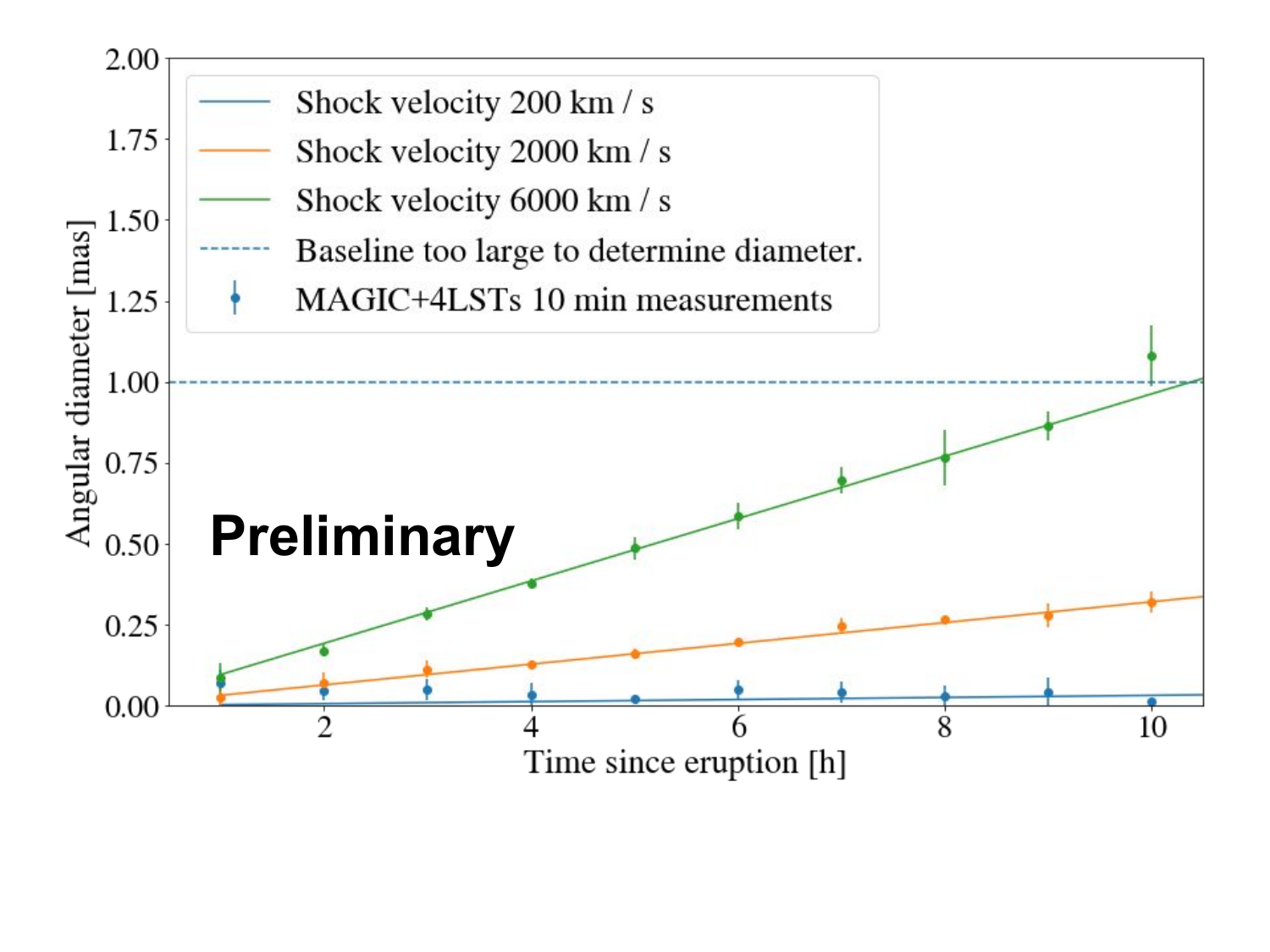}
    \caption{\textit{Simulation of an observation of a $B=5.5~mag$ nova at a distance of $3~Kpc$ with MAGIC+4CTAO-LSTs, showing the angular resolution of snapshots of 10 min. The size of a nova with a low expansion velocity could be measured for days or even for a month depending on the shock velocity. Once the angular diameter of the expanding shell is $> 1~mas$, the available baselines are too large to detect it.}}
    \label{fig:nova}
\end{figure}

MAGIC+4CTAO-LSTs will be ideal to study the first days of expansion of galactic novae, thanks to its sensitivity, angular resolution, UV plane coverage and the technical implementation that will allow to change from gamma-ray observations to SII mode in a matter of minutes. To prove this we have simulated the observation of a $B=5.5~mag$ nova at a distance of 3 kpc with MAGIC+4CTAO-LSTs taking 10min snapshots to measure the expansion rate in multiple directions (Fig. \ref{fig:nova})

\newpage

\section{Conclusions}

We have shown results of a tested solution to implement the Intensity Interferometry technique to IACTS with MAGIC SII. This approach has been extended to the CTAO-LST1 with minimal hardware modifications and now MAGIC+CTAO-LST1 SII has started operations successfully. The same modifications done to CTAO-LST1 are planed for the upcoming CTAO-LSTs, and we think MAGIC+4CTAO-LSTs SII could start operations in a few years.

We have shown the great scientific potential of MAGIC+CTAO-LST1 SII. Furthermore, this will dramatically improve once MAGIC+4CTAO-LSTs SII is operative.

Finally, we want to stress that the MAGIC+CTAO-LST1 SII shows a very realistic solution to implement SII to the whole CTAO.

\acknowledgments 
 
The MAGIC collaboration would like to thank the Instituto de Astrof\'{\i}sica de Canarias for the excellent working conditions at the Observatorio del Roque de los Muchachos in La Palma. The financial support of the German BMBF, MPG and HGF; the Italian INFN and INAF; the Swiss National Fund SNF; the grants PID2019-104114RB-C31, PID2019-104114RB-C32, PID2019-104114RB-C33, PID2019-105510GB-C31, PID2019-107847RB-C41, PID2019-107847RB-C42, PID2019-107847RB-C44, PID2019-107988GB-C22, PID2022-136828NB-C41, PID2022-137810NB-C22, PID2022-138172NB-C41, PID2022-138172NB-C42, PID2022-138172NB-C43, PID2022-139117NB-C41, PID2022-139117NB-C42, PID2022-139117NB-C43, PID2022-139117NB-C44 funded by the Spanish MCIN/AEI/ 10.13039/501100011033 and “ERDF A way of making Europe”; the Indian Department of Atomic Energy; the Japanese ICRR, the University of Tokyo, JSPS, and MEXT; the Bulgarian Ministry of Education and Science, National RI Roadmap Project DO1-400/18.12.2020 and the Academy of Finland grant nr. 320045 is gratefully acknowledged. This work was also been supported by Centros de Excelencia ``Severo Ochoa'' y Unidades ``Mar\'{\i}a de Maeztu'' program of the Spanish MCIN/AEI/ 10.13039/501100011033 (CEX2019-000920-S, CEX2019-000918-M, CEX2021-001131-S) and by the CERCA institution and grants 2021SGR00426 and 2021SGR00773 of the Generalitat de Catalunya; by the Croatian Science Foundation (HrZZ) Project IP-2022-10-4595 and the University of Rijeka Project uniri-prirod-18-48; by the Deutsche Forschungsgemeinschaft (SFB1491) and by the Lamarr-Institute for Machine Learning and Artificial Intelligence; by the Polish Ministry Of Education and Science grant No. 2021/WK/08; and by the Brazilian MCTIC, CNPq and FAPERJ.

\bigskip

The CTAO-LST project gratefully acknowledge financial support from the following agencies and organizations:

\bigskip

State Committee of Science of Armenia, Armenia;
The Australian Research Council, Astronomy Australia Ltd, The University of Adelaide, Australian National University, Monash University, The University of New South Wales, The University of Sydney, Western Sydney University, Australia; Federal Ministry of Education, Science and Research, and Innsbruck University, Austria;
Conselho Nacional de Desenvolvimento Cient\'{\i}fico e Tecnol\'{o}gico (CNPq), Funda\c{c}\~{a}o de Amparo \`{a} Pesquisa do Estado do Rio de Janeiro (FAPERJ), Funda\c{c}\~{a}o de Amparo \`{a} Pesquisa do Estado de S\~{a}o Paulo (FAPESP), Funda\c{c}\~{a}o de Apoio \`{a} Ci\^encia, Tecnologia e Inova\c{c}\~{a}o do Paran\'a - Funda\c{c}\~{a}o Arauc\'aria, Ministry of Science, Technology, Innovations and Communications (MCTIC), Brasil;
Ministry of Education and Science, National RI Roadmap Project DO1-153/28.08.2018, Bulgaria; 
The Natural Sciences and Engineering Research Council of Canada and the Canadian Space Agency, Canada; 
ANID PIA/APOYO AFB230003, ANID-Chile Basal grant FB 210003, N\'ucleo Milenio TITANs (NCN19-058), FONDECYT-Chile grants 1201582, 1210131, 1230345, and 1240904; 
Croatian Science Foundation, Rudjer Boskovic Institute, University of Osijek, University of Rijeka, University of Split, Faculty of Electrical Engineering, Mechanical Engineering and Naval Architecture, University of Zagreb, Faculty of Electrical Engineering and Computing, Croatia;
Ministry of Education, Youth and Sports, MEYS  LM2018105, LM2023047, EU/MEYS CZ.02.1.01/0.0/0.0/16\_013/0001403, CZ.02.1.01/0.0/0.0/18\_046/0016007,  CZ.02.1.01/0.0/0.0/16\_019/0000754 and CZ.02.01.01/00/22\_008/0004632, Czech Republic; 
Academy of Finland (grant nr.317636 and 320045), Finland;
Ministry of Higher Education and Research, CNRS-INSU and CNRS-IN2P3, CEA-Irfu, ANR, Regional Council Ile de France, Labex ENIGMASS, OCEVU, OSUG2020 and P2IO, France; 
The German Ministry for Education and Research (BMBF), the Max Planck Society, the German Research Foundation (DFG, with Collaborative Research Centres 876 \& 1491), and the Helmholtz Association, Germany; 
Department of Atomic Energy, Department of Science and Technology, India; 
Istituto Nazionale di Astrofisica (INAF), Istituto Nazionale di Fisica Nucleare (INFN), MIUR, Istituto Nazionale di Astrofisica (INAF-OABRERA) Grant Fondazione Cariplo/Regione Lombardia ID 2014-1980/RST\_ERC, Italy; 
ICRR, University of Tokyo, JSPS, MEXT, Japan; 
Netherlands Research School for Astronomy (NOVA), Netherlands Organization for Scientific Research (NWO), Netherlands; 
University of Oslo, Norway; 
Ministry of Science and Higher Education, DIR/WK/2017/12, the National Centre for Research and Development and the National Science Centre, UMO-2016/22/M/ST9/00583, Poland; 
Slovenian Research Agency, grants P1-0031, P1-0385, I0-0033, J1-9146, J1-1700, N1-0111, and the Young Researcher program, Slovenia; 
South African Department of Science and Technology and National Research Foundation through the South African Gamma-Ray Astronomy Programme, South Africa; 
The Spanish groups acknowledge funds from "ERDF A way of making Europe" and the Spanish Ministry of Science and Innovation and the Spanish Research State Agency (AEI) via MCIN/AEI/10.13039/501100011033 through government budget lines PGE2021/28.06.000X.411.01, PGE2022/28.06.000X.411.01, PGE2022/28.06.000X.711.04, and grants PID2022-137810NB-C22, PID2022-136828NB-C42, PID2022-139117NB-C42, PID2022-139117NB-C41, PID2022-136828NB-C41, PID2022-138172NB-C43, PID2022-138172NB-C42, PID2022-139117NB-C44, PID2021-124581OB-I00, PID2021-125331NB-I00, PID2019-104114RB-C31,  PID2019-107847RB-C44, PID2019-104114RB-C32, PID2019-105510GB-C31, PID2019-104114RB-C33, PID2019-107847RB-C41, PID2019-107847RB-C43, PID2019-107847RB-C42; the "Centro de Excelencia Severo Ochoa" program through grants no. CEX2019-000920-S, CEX2020-001007-S, CEX2021-001131-S; the "Unidad de Excelencia Mar\'ia de Maeztu" program through grants no. CEX2019-000918-M, CEX2020-001058-M; the "Ram\'on y Cajal" program through grants RYC2021-032552-I, RYC2021-032991-I, RYC2020-028639-I and RYC-2017-22665; and the "Juan de la Cierva" program through grants no. IJC2019-040315-I and JDC2022-049705-I. La Caixa Banking Foundation is also acknowledged, grant no. LCF/BQ/PI21/11830030. They also acknowledge the project "Tecnolog\'ias avanzadas para la exploraci\'on del universo y sus componentes" (PR47/21 TAU), funded by Comunidad de Madrid regional government. Funds were also granted by the Junta de Andaluc\'ia regional government under the "Plan Complementario de I+D+I" (Ref. AST22\_00001) and "Plan Andaluz de Investigaci\'on, Desarrollo e Innovaci\'on" (Ref. FQM-322); by the "Programa Operativo de Crecimiento Inteligente" FEDER 2014-2020 (Ref.~ESFRI-2017-IAC-12) and Spanish Ministry of Science and Innovation, 15\% co-financed by "Consejer\'ia de Econom\'ia, Industria, Comercio y Conocimiento" of the Gobierno de Canarias regional government. The Generalitat de Catalunya regional government is also gratefully acknowledged via its "CERCA'' program and grants 2021SGR00426 and 2021SGR00679. Spanish groups were also kindly supported by European Union funds via the "Horizon 2020" program, grant no. GA:824064, and NextGenerationEU, grants no. PRTR-C17.I1, CT19/23-INVM-109, and "Mar\'ia Zambrano" program, BDNS: 572725. This research used computing and storage resources provided by the Port d'Informaci\'o Cient\'ifica (PIC) data center; 
Swedish Research Council, Royal Physiographic Society of Lund, Royal Swedish Academy of Sciences, The Swedish National Infrastructure for Computing (SNIC) at Lunarc (Lund), Sweden; 
State Secretariat for Education, Research and Innovation (SERI) and Swiss National Science Foundation (SNSF), Switzerland; Durham University, Leverhulme Trust, Liverpool University, University of Leicester, University of Oxford, Royal Society, Science and Technology Facilities Council, UK; 
U.S. National Science Foundation, U.S. Department of Energy, Argonne National Laboratory, Barnard College, University of California, University of Chicago, Columbia University, Georgia Institute of Technology, Institute for Nuclear and Particle Astrophysics (INPAC-MRPI program), Iowa State University, the Smithsonian Institution, V.V.D. is funded by NSF grant AST-1911061, Washington University McDonnell Center for the Space Sciences, The University of Wisconsin and the Wisconsin Alumni Research Foundation, USA.

\bigskip

The research leading to these results has received funding from the European Union's Seventh Framework Programme (FP7/2007-2013) under grant agreements No~262053 and No~317446.
This project is receiving funding from the European Union's Horizon 2020 research and innovation programs under agreement No~676134.

\bigskip

Funded/Co-funded by the European Union (ERC, MicroStars, 101076533). Views and opinions expressed are however those of the author(s) only and do not necessarily reflect those of the European Union or the European Research Council. Neither the European Union nor the granting authority can be held responsible for them.

\bigskip

\bibliography{main} 
\bibliographystyle{spiebib} 

\section*{Full author list: The MAGIC Collaboration}

K.~Abe$^{1}$, 
S.~Abe$^{2}$, 
J.~Abhir$^{3}$, 
A.~Abhishek$^{4}$, 
V.~A.~Acciari$^{5}$, 
A.~Aguasca-Cabot$^{6}$, 
I.~Agudo$^{7}$, 
T.~Aniello$^{8}$, 
S.~Ansoldi$^{9,45}$, 
L.~A.~Antonelli$^{8}$
% ...
\\

\noindent \normalsize{$^{1}$ Japanese MAGIC Group: Department of Physics, Tokai University, Hiratsuka, 259-1292 Kanagawa, Japan,}
\normalsize{$^{2}$ Japanese MAGIC Group: Institute for Cosmic Ray Research (ICRR), The University of Tokyo, Kashiwa, 277-8582 Chiba, Japan,}
\normalsize{$^{3}$ ETH Z\"urich, CH-8093 Z\"urich, Switzerland,}
\normalsize{$^{4}$ Universit\`a di Siena and INFN Pisa, I-53100 Siena, Italy,}
\normalsize{$^{5}$ Institut de F\'isica d'Altes Energies (IFAE), The Barcelona Institute of Science and Technology (BIST), E-08193 Bellaterra (Barcelona), Spain,}
\normalsize{$^{6}$ Universitat de Barcelona, ICCUB, IEEC-UB, E-08028 Barcelona, Spain,}
\normalsize{$^{7}$ Instituto de Astrof\'isica de Andaluc\'ia-CSIC, Glorieta de la Astronom\'ia s/n, 18008, Granada, Spain,}
\normalsize{$^{8}$ National Institute for Astrophysics (INAF), I-00136 Rome, Italy,}
\normalsize{$^{9}$ Universit\`a di Udine and INFN Trieste, I-33100 Udine, Italy,}
\normalsize{$^{10}$ Max-Planck-Institut f\"ur Physik, D-85748 Garching, Germany,}
\normalsize{$^{11}$ Universit\`a di Padova and INFN, I-35131 Padova, Italy,}
\normalsize{$^{12}$ Croatian MAGIC Group: University of Zagreb, Faculty of Electrical Engineering and Computing (FER), 10000 Zagreb, Croatia,}
\normalsize{$^{13}$ Centro Brasileiro de Pesquisas F\'isicas (CBPF), 22290-180 URCA, Rio de Janeiro (RJ), Brazil,}
\normalsize{$^{14}$ IPARCOS Institute and EMFTEL Department, Universidad Complutense de Madrid, E-28040 Madrid, Spain,}
\normalsize{$^{15}$ Instituto de Astrof\'isica de Canarias and Dpto. de  Astrof\'isica, Universidad de La Laguna, E-38200, La Laguna, Tenerife, Spain,}
\normalsize{$^{16}$ University of Lodz, Faculty of Physics and Applied Informatics, Department of Astrophysics, 90-236 Lodz, Poland,}
\normalsize{$^{17}$ Centro de Investigaciones Energ\'eticas, Medioambientales y Tecnol\'ogicas, E-28040 Madrid, Spain,}
\normalsize{$^{18}$ Departament de F\'isica, and CERES-IEEC, Universitat Aut\`onoma de Barcelona, E-08193 Bellaterra, Spain,}
\normalsize{$^{19}$ Universit\`a di Pisa and INFN Pisa, I-56126 Pisa, Italy,}
\normalsize{$^{20}$ INFN MAGIC Group: INFN Sezione di Bari and Dipartimento Interateneo di Fisica dell'Universit\`a e del Politecnico di Bari, I-70125 Bari, Italy,}
\normalsize{$^{21}$ Armenian MAGIC Group: A. Alikhanyan National Science Laboratory, 0036 Yerevan, Armenia,}
\normalsize{$^{22}$ Department for Physics and Technology, University of Bergen, Norway,}
\normalsize{$^{23}$ INFN MAGIC Group: INFN Sezione di Torino and Universit\`a degli Studi di Torino, I-10125 Torino, Italy,}
\normalsize{$^{24}$ Croatian MAGIC Group: University of Rijeka, Faculty of Physics, 51000 Rijeka, Croatia,}
\normalsize{$^{25}$ Universit\"at W\"urzburg, D-97074 W\"urzburg, Germany,}
\normalsize{$^{26}$ Technische Universit\"at Dortmund, D-44221 Dortmund, Germany,}
\normalsize{$^{27}$ Japanese MAGIC Group: Physics Program, Graduate School of Advanced Science and Engineering, Hiroshima University, 739-8526 Hiroshima, Japan,}
\normalsize{$^{28}$ Deutsches Elektronen-Synchrotron (DESY), D-15738 Zeuthen, Germany,}
\normalsize{$^{29}$ Armenian MAGIC Group: ICRANet-Armenia, 0019 Yerevan, Armenia,}
\normalsize{$^{30}$ Croatian MAGIC Group: University of Split, Faculty of Electrical Engineering, Mechanical Engineering and Naval Architecture (FESB), 21000 Split, Croatia,}
\normalsize{$^{31}$ Croatian MAGIC Group: Josip Juraj Strossmayer University of Osijek, Department of Physics, 31000 Osijek, Croatia,}
\normalsize{$^{32}$ Finnish MAGIC Group: Finnish Centre for Astronomy with ESO, Department of Physics and Astronomy, University of Turku, FI-20014 Turku, Finland,}
\normalsize{$^{33}$ University of Geneva, Chemin d'Ecogia 16, CH-1290 Versoix, Switzerland,}
\normalsize{$^{34}$ Saha Institute of Nuclear Physics, A CI of Homi Bhabha National Institute, Kolkata 700064, West Bengal, India,}
\normalsize{$^{35}$ Inst. for Nucl. Research and Nucl. Energy, Bulgarian Academy of Sciences, BG-1784 Sofia, Bulgaria,}
\normalsize{$^{36}$ Japanese MAGIC Group: Department of Physics, Yamagata University, Yamagata 990-8560, Japan,}
\normalsize{$^{37}$ Finnish MAGIC Group: Space Physics and Astronomy Research Unit, University of Oulu, FI-90014 Oulu, Finland,}
\normalsize{$^{38}$ Japanese MAGIC Group: Chiba University, ICEHAP, 263-8522 Chiba, Japan,}
\normalsize{$^{39}$ Japanese MAGIC Group: Institute for Space-Earth Environmental Research and Kobayashi-Maskawa Institute for the Origin of Particles and the Universe, Nagoya University, 464-6801 Nagoya, Japan,}
\normalsize{$^{40}$ Croatian MAGIC Group: Rudjer Bovskovi\'c Institute, 10000 Zagreb, Croatia,}
\normalsize{$^{41}$ Japanese MAGIC Group: Department of Physics, Kyoto University, 606-8502 Kyoto, Japan,}
\normalsize{$^{42}$ INFN MAGIC Group: INFN Sezione di Perugia, I-06123 Perugia, Italy,}
\normalsize{$^{43}$ INFN MAGIC Group: INFN Roma Tor Vergata, I-00133 Roma, Italy,}
\normalsize{$^{44}$ Japanese MAGIC Group: Department of Physics, Konan University, Kobe, Hyogo 658-8501, Japan,}
\normalsize{$^{45}$ also at International Center for Relativistic Astrophysics (ICRA), Rome, Italy,}
\normalsize{$^{46}$ also at Port d'Informaci\'o Cient\'ifica (PIC), E-08193 Bellaterra (Barcelona), Spain,}
\normalsize{$^{47}$ also at Department of Physics, University of Oslo, Norway,}
\normalsize{$^{48}$ also at Dipartimento di Fisica, Universit\`a di Trieste, I-34127 Trieste, Italy,}
\normalsize{$^{49}$ Max-Planck-Institut f\"ur Physik, D-85748 Garching, Germany,}
\normalsize{$^{50}$ also at INAF Padova,}
\normalsize{$^{51}$ Japanese MAGIC Group: Institute for Cosmic Ray Research (ICRR), The University of Tokyo, Kashiwa, 277-8582 Chiba, Japan}

\section*{Full author list: The CTAO-LST Project}

{\noindent
K.~Abe$^{1}$,
S.~Abe$^{2}$,
A.~Abhishek$^{3}$,
F.~Acero$^{4,5}$,
A.~Aguasca-Cabot$^{6}$,
I.~Agudo$^{7}$,
C.~Alispach$^{8}$,
N.~Alvarez~Crespo$^{9}$,
D.~Ambrosino$^{10}$,
L.~A.~Antonelli$^{11}$
% ...
}\\

{\noindent
$^{1}${Department of Physics, Tokai University, 4-1-1, Kita-Kaname, Hiratsuka, Kanagawa 259-1292, Japan}.\\
$^{2}${Institute for Cosmic Ray Research, University of Tokyo, 5-1-5, Kashiwa-no-ha, Kashiwa, Chiba 277-8582, Japan}.
$^{3}${INFN and Università degli Studi di Siena, Dipartimento di Scienze Fisiche, della Terra e dell'Ambiente (DSFTA), Sezione di Fisica, Via Roma 56, 53100 Siena, Italy}.
$^{4}${Université Paris-Saclay, Université Paris Cité, CEA, CNRS, AIM, F-91191 Gif-sur-Yvette Cedex, France}.
$^{5}${FSLAC IRL 2009, CNRS/IAC, La Laguna, Tenerife, Spain}.
$^{6}${Departament de Física Quàntica i Astrofísica, Institut de Ciències del Cosmos, Universitat de Barcelona, IEEC-UB, Martí i Franquès, 1, 08028, Barcelona, Spain}.
$^{7}${Instituto de Astrofísica de Andalucía-CSIC, Glorieta de la Astronomía s/n, 18008, Granada, Spain}.
$^{8}${Department of Astronomy, University of Geneva, Chemin d'Ecogia 16, CH-1290 Versoix, Switzerland}.
$^{9}${IPARCOS-UCM, Instituto de Física de Partículas y del Cosmos, and EMFTEL Department, Universidad Complutense de Madrid, Plaza de Ciencias, 1. Ciudad Universitaria, 28040 Madrid, Spain}.
$^{10}${INFN Sezione di Napoli, Via Cintia, ed. G, 80126 Napoli, Italy}.
$^{11}${INAF - Osservatorio Astronomico di Roma, Via di Frascati 33, 00040, Monteporzio Catone, Italy}.
% ...
}

\end{document}